\pdfoutput=1%&pdflatex
\documentclass[letterpaper,twocolumn,amsmath,amssymb,rmp,twoside]{revtex4}
\usepackage{amssymb,amsmath}
\usepackage{mathtime}

\usepackage{color}
\definecolor{darkblue}{rgb}{0,0,0.5}
\usepackage[a4paper,pdfpagemode=None,colorlinks,plainpages=false,pdfpagelabels,bookmarksnumbered,citecolor=darkblue,linkcolor=darkblue]{hyperref}

\usepackage{graphicx}
\usepackage[left=1.5cm,right=2cm,top=2cm,bottom=2.5cm]{geometry}

\usepackage{url}

\usepackage{pstricks,pst-node,pst-tree}

\usepackage[cmtip]{xypic}
\xyoption{all}
\usepackage{array}

\DeclareMathOperator{\Prob}{Prob} \DeclareMathOperator{\Pois}{Pois}
 \DeclareMathOperator{\Exp}{Exp}

\newcommand{\EE}{\mathbb{E}}
\newcommand{\II}{\mathbb{I}}

\newcommand{\pa}{\mathrm{pa}}
\newcommand{\Exit}{\mathrm{Exit}}

\newcommand{\rmd}{\,\mathrm{d}}

\begin{document}

\author{Moritz Gerstung}
\email{moritz.gerstung@bsse.ethz.ch}
\author{Niko Beerenwinkel}
\email{niko.beerenwinkel@bsse.ethz.ch}

\affiliation{Department of Biosystems Science and Engineering, ETH Zurich,
Mattenstrasse 26, 4058 Basel, Switzerland}

% \begin{center}

% \vspace*{2cm}

% {\huge \bf Waiting time models of cancer progression}

% \vspace{1cm}

% (Running head: Cancer progression)

% \vspace{2cm}

% {\large \bf Moritz Gerstung and Niko Beerenwinkel}

% \vspace{2cm}

% Department of Biosystems Science and Engineering\\ ETH Zurich\\
% Mattenstrasse 26\\ 4058 Basel\\ Switzerland\\[1ex]
% \texttt{moritz.gerstung@bsse.ethz.ch}\\[1ex]
% \texttt{niko.beerenwinkel@bsse.ethz.ch}
% \end{center}

% \vfill \hfill \today

%\newpage

\title[Cancer progression]{Waiting time models of cancer progression}

%\begin{document}

%\author[N. Beerenwinkel]{Niko Beerenwinkel}
%\address{Department of Biosystems Science and Engineering,
% Mattenstrasse 26, 4058 Basel, Switzerland}
%\email{niko.beerenwinkel@bsse.ethz.ch}

\begin{abstract}
  Cancer progression is an evolutionary process that is driven by
  mutation and selection in a population of tumor cells.  We discuss
  mathematical models of cancer progression, starting from traditional
  multistage theory.  Each stage is associated with the occurrence of
  genetic alterations and their fixation in the population.  We
  describe the accumulation of mutations using conjunctive Bayesian
  networks, an exponential family of waiting time models in which the
  occurrence of mutations is constrained to a partial temporal order.
  Two opposing limit cases arise if mutations either follow a linear
  order or occur independently.  We derive exact analytical
  expressions for the waiting time until a specific number of
  mutations have accumulated in these limit cases as well as for the
  general conjunctive Bayesian network.
  % and show how these relate to the dependencies between the
  % mutations.
  Finally, we analyze a stochastic population genetics model that
  explicitly accounts for mutation and selection.  In this model,
  waves of clonal expansions sweep through the population at
  equidistant intervals.  We present an approximate analytical
  expression for the waiting time in this model and compare it to the
  results obtained for the conjunctive Bayesian networks.
\\[1.5ex]
%\footnotesize 
\noindent {\bf Keywords:}
{Bayesian network, cancer, genetic progression, multistage theory,
Wright-Fisher process}
\end{abstract}
%\begin{quote}

%\end{quote}

\maketitle

\section{Introduction}

Cancer is a genetic disease that develops as the result of mutations
in specific genes.
%, or functional alterations in
%DNA regions containing these genes \citep{Vogelstein2004}.
When these genes work normally, they control the growth of cells in
the body. Cancer cells have lost the normal cooperative behavior of
cells in multicellular organisms resulting in increased
proliferation. Tumor development starts from a single genetically
altered cell and proceeds by successive clonal expansions of cells
that have acquired additional advantageous mutations. The progression
of cancer is characterized by the accumulation of these genetic
changes
\citep{CairnsN1975,NowellS1976,MerloNRC2006,Michor2004,Crespi2005}.

Many oncogenes and tumor suppressor genes have been identified that
contribute to tumorigenesis \citep{FutrealNRC2004}. In general, the
mutational patterns of cancer cells vary greatly, not only among
cancer types, but also among individual tumors of the same type. Some
of this genetic variation might be due to the fact that all cancer
cells need to acquire certain functional changes, the hallmarks of
cancer, and most of these functions are accomplished by several gene
products acting together in signaling pathways
\citep{Hanahan2000,Vogelstein2004}. Thus, many different genetic
alterations can have similar phenotypic effects.

The incidence of sporadic cancer indicates that the underlying
events are stochastic and that, in general, several steps are necessary.
Therefore, a random processes approach appears
to be an appropriate modeling strategy.
The progression stages are generally not observable on a molecular level
{\em in vivo}, and in a clinical setting, patients are typically diagnosed at
the final stages of tumorigenesis.
%It therefore remains unknown
%whether the emergent properties from a combination of the single steps
%are sufficient to account for the observed bulk phenomena.
Mathematical modeling plays an important role in
cancer research today, because it can be used to reconstruct and to analyze
the evolutionary process driving cancer progression
\citep{Anderson2008}.

Models of tumorigenesis have been proposed early on to explain cancer
incidence data \citep{Nordling1953,Armitage1954,Knudson1971}. These
models assume that cancer is a stochastic multistep process with small
transition rates and they have been further developed into the
multistage theory of cancer
\citep{Moolgavkar1992,Frank2007,Jones2008}. The tumor stages may be
defined by specific mutations, by the number of mutations, by
epigenetic changes, by functional alterations, or by histological
properties. Since cancer progression is an evolutionary process,
population genetics models are used extensively to describe
tumorigenesis
\citep{Nowak2006,Wodarz2005,Schinazi2006,Beerenwinkel2007c,Durrett2008}.
Various deterministic and stochastic models have been proposed, some
of which address specific questions, such as the dynamics of tumor
suppressor genes \citep{Iwasa2005}, genetic instability
\citep{Nowak2006b}, or tissue architecture \citep{Nowak2003}.

As more and more genetic data from cancer cells become available from
comprehensive studies
\citep{SjoblomS2006,WoodS2007,JonesS2008,ParsonsS2008,LeyN2008} and through databases
\citep{FutrealNRC2004,Baudis2007,MitelmanDB2008}, one can
also start investigating the dependencies between genetic events using
statistical models. In view of multistage theory, tumors
proceed through distinct stages, which can be characterized by the appearance
of certain mutations. Particular attention
has been paid to inferring the order of genetic alterations. Several
graphical models have been developed for this purpose and applied to
various cancer types
\citep{Desper1999,Radmacher2001,Hjelm2006,Rahnenfuehrer2005,
  Beerenwinkel2006a,Beerenwinkel2007d,Beerenwinkel2007e}.

A quantitative understanding of carcinogenesis can help developing new
diagnostic and prognostic markers.  Today a variety of univariate
genetic markers is known \citep{SidranskyNRC2002}, most of them
comprising well-known oncogenes or tumor suppressors. Because of the
diverse genetic nature of cancer, markers measuring the accumulation
of several mutations, i.e., the progression of cancer, may improve
existing ones.  Here, we investigate the dynamics of cancer progression
as a function of transition rates and of order constraints on the
genetic events. The expected waiting time can be regarded as a measure
of genetic progression to cancer \citep{Rahnenfuehrer2005}.

In Section~\ref{sec:multistage-theory}, we introduce the general
stochastic multistep process and present an equivalent description in
terms of ordinary differential equations (ODEs).  At this abstract
level of description, carcinogenesis may be regarded as proceeding
through distinct stages, which can be defined by histological grades,
functional changes, or genetic alterations.  In
Section~\ref{sec:genet-progr-canc}, these stages will be associated
with the occurrence of a certain number of mutations.  We present
expressions of the waiting time until a given stage is reached, for
different models of mutation.  Finally, in
Section~\ref{sec:population-dynamics}, we analyze an evolutionary
model of carcinogenesis explicitly describing the appearance of
genetic alterations in the tissue by mutations in single cells and
their subsequent clonal expansions.

\section{Multistage theory}
\label{sec:multistage-theory}

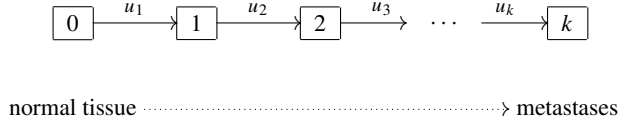
\begin{figure}
\centering
\begin{tabular}{c}
\xymatrix@!C=0.8cm@R=0.9cm{
    *+[F]{~0~}\ar@{->}[r]^{u_1} &
    *+[F]{~1~}\ar@{->}[r]^{u_2} &
    *+[F]{~2~}\ar@{->}[r]^{u_3} &
    ~~\dots~~\ar@{->}[r]^{u_k} &
    *+[F]{~k~}\\
    \smash{\text{normal tissue}} \ar@<.5ex>@{.>}[rrrr]&&&& \smash{\text{metastases}}
}
\end{tabular}
\caption{\label{fig:multistage}
The multistage model. Cancer is assumed to develop in a linear process
consisting of $k$ stages. The progression steps mark the major transitions
of the tissue from normal to cancerous and eventually to metastasizing.
At each transitions $(j-1) \to j$ the waiting time follows an exponential
distribution with parameter $u_j$ giving rise to the model defined by
Eq.~(\ref{eq:tau}).
}
\end{figure}

The multistage theory of cancer postulates that tumorigenesis is a
linear multistep process, in which each step from one stage to the
next is a rare event (Figure~\ref{fig:multistage}).  Let us denote the
cancer stages by $0$, $1$, $2$, $\dots$, $k$, where stage $0$ refers
to the normal precancerous state, $1$ to the first adenomatous stage,
and $k$ to a defined cancerous endpoint, such as the the formation of
metastases.  The process is started at time $t=0$ in state $0$.

The transition rate from stage $j-1$ to stage $j$ is denoted
$u_j$. That is, the waiting times for the transitions to occur are
assumed to be independently exponentially distributed. Here, the coefficients $u_j$ denote the transition
rates between the stages of tumor development; later we will link them
to different models of mutation and fixation.  Because of
the sequential nature of the linear model, the waiting time $\tau_k$
until stage $k$ is reached is given recursively by the sum of
exponentially distributed random variables,
\begin{equation}  \label{eq:tau}
 \tau_1 \sim \Exp(u_1), \qquad  \tau_j \sim \tau_{j-1} + \Exp(u_j), \quad j = 2,\dots, k.
\end{equation}
The waiting times follow the linear order
$\tau_1 < \dots < \tau_k$ and the expected waiting
time is
\begin{equation}
  \label{eq:waitingtimelinear}
    \EE[\tau_k] = \EE[\tau_{k-1}] + \frac{1}{u_k}
        = \sum_{j=1}^k \frac{1}{u_j}.
\end{equation}
In particular, if all transition rates are equal, $u_j = u$ for all
$j=1, \dots, k$, we find $\EE[\tau_k] = k/u$. Hence the waiting time
scales linear in the number of transitions $k$.

Let $f_{\tau_1,\dots,\tau_k}(t_1, \dots, t_k)$ be the density function of the joint
distribution of waiting times $\tau = (\tau_1, \dots, \tau_k)$ defined
by Eq.~(\ref{eq:tau}).  The linear order of the waiting times $\tau_j$
induces the factorization
\begin{equation}  \label{eq:lineardensity}
    f_{\tau_1,\dots,\tau_k}(t_1, \dots, t_k) = \prod_{j=1}^k f_{\tau_j|\tau_{j-1}}(t_j \mid t_{j-1})
\end{equation}
of $f_{\tau_1,\dots,\tau_k}$ into the conditional densities
\begin{equation}
  \begin{split}
    \label{eq:transition}
    f_{\tau_j|\tau_{j-1}}(t_j \mid t_{j-1}) = u_j \exp \left( -u_j[t_j -
      t_{j-1}] \right) \, \II(t_j > t_{j-1}),
  \end{split}
\end{equation}
where $\II$ is the indicator function.

%\subsection{ODE}
Multistage theory can also be formulated as a system of ordinary
differential equations (ODEs). We derive this formulation as follows:
Let $x_j(t)$ denote the probability that stage $j$ is reached before
time $t \ge 0$, but stage $j+1$ has not yet been reached,
\begin{eqnarray}
    x_0(t) &=& \Prob[0 < t < \tau_1] \nonumber \\
    x_j(t) &=& \Prob[\tau_j < t < \tau_{j+1}],
        \quad  j = 1, \dots k-1, \\
    x_k(t) &=& \Prob[\tau_k < t]. \nonumber
\end{eqnarray}
We have $x_0(t) + \dots + x_k(t) = 1$ and $x_j(t) = \Prob[t <
\tau_{j+1}] - \Prob[t < \tau_j]$ due to the linearity of transitions.
It follows that $\dot{x}_j(t) = f_{\tau_{j+1}}(t) - f_{\tau_j}(t)$.% , because
% $\Prob(\tau_k < t)$ denotes a cumulative waiting time
% distribution.
Using the conditional exponential nature of the model,
Eq.~(\ref{eq:transition}), one finds that $f_{\tau_j}(t) = \int_0^\infty
f_{\tau_j,\tau_{j-1}}(t, t') \rmd t' =\int_0^{t} \exp(-u_{j}[t-t'])
f_{\tau_{j-1}}(t') \rmd t' $. From the identity
$\exp(-u_{j}t)=u_{j}\int_{t}^\infty \exp(-u_{j}t') \rmd t'$, one obtains
\begin{align}\nonumber 
  f_{\tau_j}(t) &= u_{j} \int_{t}^\infty\!\!\! \int_0^{t}\exp(-u_{j}[t''-t'])
  f_{\tau_{j-1}}(t') \rmd t' \rmd t''\\
  & = u_{j}
  \int_{t}^\infty\!\!\! \int_0^{t}f_{\tau_j,\tau_{j-1}}(t'',t') \rmd t' \rmd t''\\
  \nonumber 
  &= u_{j}\Prob[\tau_{j-1} < t < \tau_{j}] = u_{j}
  x_{j-1}(t).
\end{align}
Hence, the probabilities $x_j(t)$ obey the set of ODEs,
\begin{eqnarray}   \label{eq:ode}
    \dot{x}_0(t) & = & -u_1 \, x_0(t), \nonumber \\
    \dot{x}_j(t) & = & u_{j} x_{j-1}(t) - u_{j+1}x_j(t),
        \quad j = 1, \dots, k-1, \quad\\
    \dot{x}_k(t) & = & u_{k} \, x_{k-1}(t), \nonumber
\end{eqnarray}
subject to initial conditions $x_0(0) = 1$ and $x_j(0)
= 0$ for all $j \ge 1$. These rate equations describe
the linear chain of exponential waiting time
processes as a probability flux of rate
$u_{j} x_{j-1}(t)$ from state $j-1$ to state $j$.

If all rates are identical, $u_j=u$ for all $j$, then the
solution of this linear system of ODEs is
given by Poisson distributions with time-dependent parameter
$ut$,
\begin{equation}
\label{eq:Pois}
    x_j(t) = \Pois(j; ut) = \frac{(ut)^j \exp(-ut)}{j!},
    \quad j = 0, \dots, k-1.
\end{equation}
The probability of having reached the final stage of
progression, $k$, at time $t$ %, $x_k(t)=\Prob[t>\tau_k]$, 
is
\begin{equation}
\label{eq:x_k}
    x_k(t) = 1 - e^{-ut} \, \sum_{j=0}^{k-1}
    \frac{(ut)^j}{j!}=\Pois(k; ut)\sum_{j=0}^\infty
\frac{(u t)^j}{(k+j)_j}.
\end{equation}
We also recover from the ODE system the expected waiting
time to the final cancer stage,
\begin{equation}
    \EE[\tau_k] = \int_0^\infty u t x_{k-1}(t) \rmd t
    = \frac{k}{u}.
\end{equation}

Multistage theory provides a mathematical framework for describing the
stepwise progression of cancer. For the above discussion of the model,
we have neither specified the definition of the postulated stages, nor the
nature of the transitions. Indeed, different interpretations and
uses of the model are possible. In the following we link multistage
theory closely to the genetic progression of cancer.

%Since there exist an absorbing boundary at $j=d$, the
%probability of finding $n$ mutations is given by
%$x_n(t)=\sum_{j=d}^\infty \Pois(j;\lambda(t))$.  Eq.~(\ref{eq:9})
%defines the probability of finding \emph{exactly} $j$ mutations at
%time $t$. The probability of finding \emph{at least} $k$ mutations is
%given by $\Prob[T_k<t]=\sum_{j\ge k}x_j(t)$. This is the cumulative
%waiting time distribution of $\tau_k=T_k$. Inserting Eq.~(\ref{eq:9})
%we find:
%\begin{equation}
%  \label{eq:4} \Prob[\tau_k<t]= x_k(t)\sum_{j=0}^\infty
%\frac{(ut)^j}{(k+j)_j}.
%% \int_0^t\Pois(k-1;ut') \rmd t' =
%% \exp(-ut)\frac{(ut)^k}{k!}\sum_{j=0}^\infty \frac{(ut)^j}{(k+j)_j}.
%\end{equation} This will be useful in the next section where we
%analyze the distribution of waiting times in a population of $N$
%identical cells.
%
%As stated above, the corresponding waiting time distribution is given
%by $f(t_k)=u{x}_{k-1}$. Thus the waiting time distribution of finding
%at least $k$ mutations is given by the mutation rate $u$ times the
%probability of finding $k-1$ mutations at time $t$. Of course, this
%simply recapitulates the conditional nature of the linear chain of
%mutations.

%Having found an explicit formula for the waiting time distribution
%allows computing the expected waiting time,
%This reproduces the expected waiting time that has been
%derived in the framework of conjunctive Bayesian networks.  In the next section we will show that
%it is extremely useful to model an ensemble of $N$ cells each obeying
%the waiting time process.

\section{Genetic progression of cancer}
\label{sec:genet-progr-canc}

In this section, we associate the stages of tumorigenesis to mutations in
the genomes of cancer cells.  Each stage is defined by the number of
mutations that have accumulated in the cells of the tissue.  Each
mutation occurs initially in a single cell as the result of an
erroneous DNA duplication.  Some mutations alter the behavior of the
cell in such a way that it experiences a growth advantage relative to
the other cells in the tissue. These cells can outgrow their
competitors in a clonal expansion and the mutation spreads in the
tissue.  While the first appearance of a mutation is essentially a
random process, i.e., each mutation is equally likely to appear, the
fate of a mutation in the population depends on the relative fitness of the
cell in which it occurs.

We define tumor progression to be in stage $j$ of the multistep model,
Eq.~(\ref{eq:tau}), if most of the tumor cells harbor exactly $j$
mutations. We are interested in the waiting time until $k$ out of $d$
possible mutations have accumulated, where typically $k \ll d$. For
example, \cite{SjoblomS2006} suggest that $k \approx 20$ genes out of
$d \approx 100$ to $1000$ need to be hit in order to develop invasive
colon cancer. In this interpretation of multistage theory, stages
correspond to population states and transitions correspond to genetic
transformations of the ensemble of tumor cells, including mutation and
selection.  These population dynamics will be investigated in more
detail in the next section. The focus of the present section is on
how different models of accumulating mutations affect the waiting
time.

%From the molecular perspective, there exist no interdependencies
%between mutations of the genome since the probability of producing an error
%at DNA duplication is more or less constant in the genome. Furthermore
%it is thought that certain mutations present growth advantages to the
%cells and their offspring. These consequently outcompete the other
%cells by their increased duplication rate. At this stage we assume
%that an (selectively advantageous) mutation will immediately fixate in
%the ensemble of cells on the time-scale of the mutation.

% Carcinogenesis is thought to require $k\approx 20$ genetic
% alterations, which confer a selective advantage.  Some of these
% mutations may occur in a random order and some may be required to
% occur in a particular order.

Genetic mutations occur randomly at erroneous cell divisions, but the
subsequent fixation of the mutation within the cell population is
 constrained: A mutation will only spread if it confers a growth
 advantage. This restricts not only the mutations driving cancer, but also the order in
which they can appear, because some physiological changes must be
achieved before others. For example, in colon cancer, cells must lose
the ability to undergo apoptosis, before additional mutations
accumulate in the resulting neoplasia. Hence loss of function of the
tumor suppressor gene {\it APC} controlling apoptosis is necessary
before other mutations such as {\it KRAS2} can fixate
\citep{Nowak2003,Vogelstein2004}. Furthermore, sometimes only the
combined action of mutations drives cancer progression.  It is known,
for example, that only the combination of {\it p53} and {\it Ras}
trigger the development of tumors in mice \citep{LandS1983}.

\begin{figure} \centering
\begin{tabular}{b{0.5cm}<{\vskip1cm}b{4.5cm}b{3cm}} (a)  &
\xymatrix@!=0.8cm{ %*+[o][F]{~0~}\ar@{->}[r]^{u_1} &
*+[o][F]{~1~}\ar@{->}[r]%^{\lambda_2}
& *+[o][F]{~2~}\ar@{->}[r]%^{\lambda_3}
%&
%~~\dots~~\ar@{->}[r]%^{\lambda_d}
%& *+[o][F]{~d~} }
& *+[o][F]{~3~} }
&
\small
\xymatrix@!=0cm{
& \bullet \ar@{..}[rr]& & \{1,2,3\} \\
\bullet\ar@{..}[rr]\ar@{..}[ur] && \bullet \ar@{..}[ur] \\
& \bullet\ar@{..}'[r][rr]\ar@{..}'[u][uu] & & \{1,2\} \ar[uu] \\
\emptyset\ar[rr]\ar@{..}[uu]\ar@{..}[ur] && \{1\}\ar@{..}[uu]\ar[ur]
}
\\%[5ex]
(b) &
\xymatrix@!=0.8cm{% &&&
%*+[o][F]{~0~}\ar@{->}[dl]^{\lambda_2}\ar@{->}[dlll]^{\lambda_1}\ar@{->}[dr]_{\lambda_j}
%\ar@{->}[drrr]_{\lambda_d} \\
*+[o][F]{~1~} & *+[o][F]{~2~} &%&\!\!\dots\!\!&&
 *+[o][F]{~3~} }
&
\small
\xymatrix@!=0cm{
& \{2,3\} \ar[rr]& & \{1,2,3\} \\
\{3\}\ar[rr]\ar[ur] && \{1,3\} \ar[ur] \\
& \{2\}\ar'[r][rr]\ar'[u][uu] & & \{1,2\} \ar[uu] \\
\emptyset\ar[rr]\ar[uu]\ar[ur] && \{1\}\ar[uu]\ar[ur]
}
\\%[5ex]
(c)  &
\xymatrix@R=0.8cm@C=.2cm{ %&
%*+[o][F]{~0~}\ar@{->}[dl]^{\lambda_1}\ar@{->}[dr]_{\lambda_2} & \\
*+[o][F]{~1~}\ar@{->}[dr]%^{\lambda_3}
& &
*+[o][F]{~2~}\ar@{->}[dl]%^{\lambda_3}
%\ar@{->}[d]%^{\lambda_4}
\\
&*+[o][F]{~3~} &%& *+[o][F]{~4~}
}
&\small
\xymatrix@!=0cm{
& \bullet \ar@{..}[rr]& & \{1,2,3\} \\
\bullet\ar@{..}[rr]\ar@{..}[ur] && \bullet \ar@{..}[ur] \\
& \{2\}\ar'[r][rr]\ar@{..}'[u][uu] & & \{1,2\} \ar[uu] \\
\emptyset\ar[rr]\ar@{..}[uu]\ar[ur] && \{1\}\ar@{..}[uu]\ar[ur]
} \\
& Poset & Genotype lattice
\end{tabular}
% \xymatrix@!=0.1cm{
% & \bullet & & \bullet \\
% \bullet && \bullet  \\
% & \bullet & & \bullet \ar[uu] \\
% \bullet\ar[rr] && \bullet\ar[ur]
% }\\
% \xymatrix@=0.3cm{
% & \bullet \ar[rr]& & \bullet \\
% \bullet\ar[rr]\ar[ur] && \bullet \ar[ur] \\
% & \bullet\ar[rr]\ar[uu] & & \bullet \ar[uu] \\
% \bullet\ar[rr]\ar[uu]\ar[ur] && \bullet\ar[uu]\ar[ur]
% }\\

\caption{\label{fig:graph} Conjunctive Bayesian networks.  Displayed
  are the Hasse diagrams of the posets representing the graph of the
  underlying Bayesian network (left) and their corresponding genotype
  lattices (right) for $d=3$ mutational events.  In the Hasse
  diagrams, each directed edge $i \to j$ denotes a direct dependency
  between two mutational events $i \prec j$; each node $j$ is
  associated with an exponential waiting time process with parameter
  $\lambda_j$ conditioned on the prevalence of all mutations with
  directed edges to $j$. The genotype lattice is the lattice of order
  ideals of the poset. It consists of all genotypes that are
  compatible with the relations of the poset. A directed edge $S \to
  T$ is drawn between two genotypes $S$, $T \subset [d]$, if $T$
  arises from $S$ by a mutation $j \in T$ and if the transition from
  $S$ to $T$ is consistent with the poset, i.e. there exists no $i
  \prec j$ with $i \not\in S$. Mutations accumulate along the
  different paths from $\emptyset$ to $[d]$ in the genotype lattice.
  Three network topologies are shown: a linear chain of mutations (a),
  independent mutations (b), and a poset in which both mutation 1 and
  mutation 2 need to occur before mutation 3 (c).  These posets induce
  different genotype lattices, namely a single path (a), the complete
  $d$-dimensional hypercube (b), and a lattice of intermediate size
  (c), respectively.  }
\end{figure}
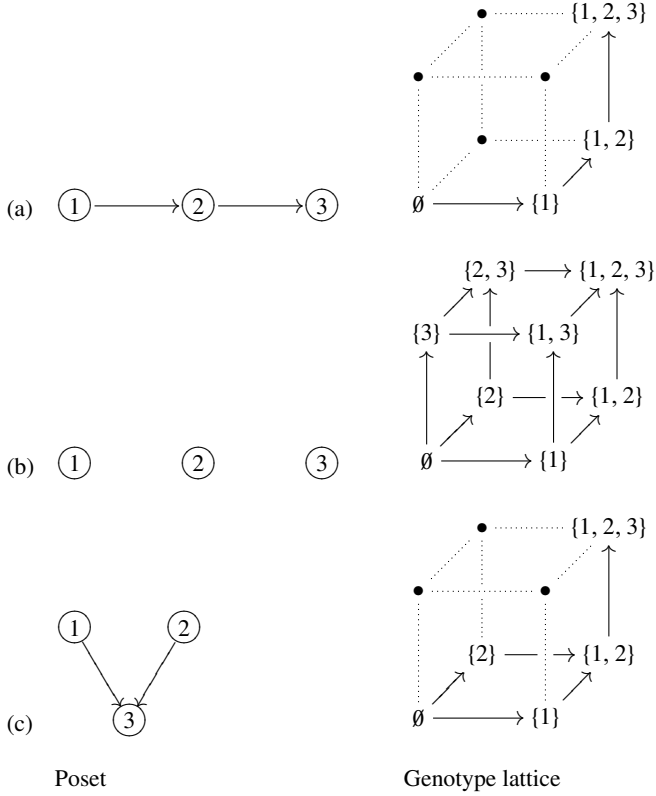

In general, there may exist several order constraints for the
successive fixation of mutations as shown in
Figure~\ref{fig:graph}. The simplest of these constraints is the
linear model, where the waiting times of all mutations are totally
ordered (Figure~\ref{fig:graph}(a)).  The linear model is exactly the
multistep process discussed in the previous section.  Alternatively,
mutations may occur independently without any constraints
(Figure~\ref{fig:graph}(b)). If there exist order relations among some
of the possible mutations, a partial order may be used to describe the
process of accumulating mutations (Figure~\ref{fig:graph}(c)).  We
will discuss these models separately and show how the topology of the
genotype space affects the waiting time.

% We will now analyze the waiting times until the appearance of certain
% mutations. The stage of carcinogenesis can then be defined by the
% occurrence of a given number of mutations. For simplicity, we assume
% that stage $k$ is reached if any $k$ mutations have accumulated.

Let $d$ be the number of possible mutations and denote by $T_j$ the
waiting time for mutation $j$ ($j=1,\dots,d$) to be generated and to
establish in the tumor. The waiting times $T_j$ are assumed to be
exponentially distributed with parameters $\lambda_j$, and they obey
certain temporal order constraints (Figure~\ref{fig:graph}). The joint
distribution of $T = (T_1, \dots, T_d)$ determines how long it takes
until $k$ out of the $d$ mutations have accumulated. We define the
random variable $\tau_k$ denoting stage $k$ as the waiting time until
any $k$ mutations appear,
\begin{equation}  \label{eq:minmax}
    \tau_k = \min_{ \{j_1,\dots,j_k\} \subset [d]} \,
        \max \, \{ T_{j_1}, \dots T_{j_k} \}, \quad [d]=\{1,2,\ldots,d\}.
\end{equation}
If all mutations accumulate in a linear order (Figure~\ref{fig:graph}(a)),
each at rate $\lambda_j$,
\begin{equation}  \label{eq:linearmutations}
    T_1 \sim \Exp(\lambda_1), \quad
    T_j \sim T_{j-1} + \Exp(\lambda_j), \quad j=2, \dots, d,
\end{equation}
then $\tau_k = T_k$ and the process of mutation and clonal expansion
is mathematically equivalent to the general linear multistep process
of Eq.~(\ref{eq:tau}). In this case, the transition rates $u_j =
\lambda_j$ may be interpreted as an effective rate for the mutation
and the clonal expansion process. According to
Eq.~(\ref{eq:waitingtimelinear}) the waiting time for $k < d$
mutations is given by $\EE[\tau_k]=\sum_{j=1}^k 1/\lambda_j$ and the
waiting time scales linear with the number of mutations.

\subsection{Independent mutations}

Let us now consider the situation where all mutations may occur in an
arbitrary order (Figure~\ref{fig:graph}(b)),
\begin{equation}  \label{eq:indepmodel}
    T_j \sim \Exp(\lambda_j), \quad j=1, \dots, d.
\end{equation}
If $k=1$, Eq.~(\ref{eq:minmax}) simplifies to
\begin{equation}  \label{eq:tau1}
    \tau_1 = \min \{T_1, \dots, T_d\}
        \sim \Exp(\lambda_1 + \dots + \lambda_d)
\end{equation}
and the expected waiting time is $\EE[\tau_1] = 1 / \sum_{j=1}^d
\lambda_j$.  If all fixation rates are equal to
$\lambda$, then $\EE[\tau_1] = 1 / (d \lambda)$. Thus, the occurrence
of any one out of $d$ mutations is equivalent to a 1-step process at
rate $u_1 = d \lambda$.

For now, we continue assuming identical rates $\lambda$. If $k \ge 2$
and the first mutation has occurred, then there are $d-1$ choices left
for the second mutation to occur, hence $\tau_2 \sim \tau_1 +
\Exp((d-1)\lambda)$.  In general, the accumulation of $k$ out of $d$
mutations, which occur independently at the same rate $\lambda$, is
equivalent to the $k$-step process, Eq.~(\ref{eq:tau}),
%The joint density of $T=(T_1,\dots,T_d)$ then reads
%\begin{equation}
%  \label{eq:densityiid}
%  f(t_1,\dots,t_d)= u^d \exp \left( -u \sum_{j=1}^d t_j \right)
%\end{equation}
%Because of the diverse nature of mutational patterns occurring in
%cancer, often the total number of mutations is the quantity of
%interest. We define the waiting times
%\begin{equation}
%  \label{eq:10}
%  \tau_k = T_{\sigma_{k}},
%\end{equation}
%where $\sigma \in S_d$ is defined to be the permutation generating a
%time-ordering of the set of waiting times, i.e., $T_{\sigma_{j}} <
%T_{\sigma_{j+1}}$ $\forall j$.
%Since there exist $d$
%possible loci for the first mutation, $\tau_1$ is exponentially
%distributed with parameter $u_1=du$. Generalizing this argument, we
%have $u_j=(d-j)u$, and hence:
%\begin{equation}  \label{eq:iid}
%    \tau_j \sim \tau_{j-1} + \Exp[(d-j+1)\lambda],
%\end{equation}
with rates $u_j = (d-j+1)\lambda$.  From Eq.~(\ref{eq:waitingtimelinear}),
we find
\begin{equation}
  \label{eq:waitingtimeiid1}
  \EE[\tau_k]=\frac{1}{\lambda}\sum_{j=1}^k\frac{1}{d-j+1}.
\end{equation}
If many mutations are possible, $d\gg k$, then the expected waiting
time for $k$ mutations is approximately $\EE[\tau_k] \approx k /(d
\lambda)$, which is smaller than $1/\lambda$ and linear in $k$. The
waiting time approaches zero in
the limit $d \rightarrow \infty$
for every fixed $k$, because the exponential distribution is non-zero
at $t = 0$. On the other hand, if $k=d$, all possible mutations need
to occur and $\EE[\tau_k] = H_k / \lambda$, where $H_k = \sum_{j=1}^k
1/j$ is the $k$-th harmonic number.  Using ${H_k}\approx {\gamma +
  \log k}$, $\gamma\approx0.577$ being the Euler-Mascheroni constant,
we find an approximate logarithmic dependency for the occurrence of
all possible mutations, $\EE[\tau_k] \approx (\gamma + \log
k)/\lambda$.  In contrast to the $k \ll d$ case, for $k = d$, the
expectation of $\tau_k$ is larger than $1/\lambda$ and increases only
logarithmically in $k$.

In both cases the expected waiting time is always larger if mutations
can only occur in a linear fashion, Eq.~(\ref{eq:waitingtimelinear}),
than if mutations are independent, Eq.~(\ref{eq:waitingtimeiid1}).
This is due to the fact that in the independent case, all mutations
are possible in any step of the process, whereas in the linear case,
only one mutation is feasible at each stage.

The ODE system, Eq.~(\ref{eq:ode}), corresponding to the multistage
model with unequal transition rates $u_j = (d-j+1)\lambda$ has also an
analytical solution.  For $j=1,\dots,d$,
\begin{equation}  \label{eq:odesolutionunequal}
    x_j(t)=\binom{d}{j} \left( 1-e^{-ut} \right)^j
    \left( e^{-ut} \right)^{d-j}.
\end{equation}
If $k \ll d$,
then $u_j \approx d \lambda$ and Eq.~(\ref{eq:Pois}) yields
%the asymptotic behavior is
%given by $e^{-d \lambda t}$. On the time-scale $1/lambda$ we may thus approximate
%$(1-e^{-\lambda t})^j\approx(\lambda t)^t$ yielding
$x_j(t)\approx \Pois(j; d \lambda t)$.
%The for a large set of mutable loci, the probability of
%finding any $j$ mutations is identical to the linear case,
%Eq.~(\ref{eq:9}), however, with a time-dependent parameter being
%$d\cdot u t$.
Thus, the number of independent mutations accumulates at a speed that is
roughly $d$ times faster than for a linear chain of mutations.

%\subsection{Independent unequal mutations}
%\label{sec:independent-case}

We now turn to the case of independent mutations with arbitrary
fixation rates $\lambda_j$.
%In this situation, the joint density
%factorizes into a product of exponentials,
%\begin{equation}
%  f(t)=\prod_{j=1}^du_j\exp(-u_jt_j).
%  \label{eq:distindep}
%\end{equation} This allows, in principle, to calculate the waiting time for a
%specific genetic pattern.
The distribution of $\tau_1$ is given by Eq.~(\ref{eq:tau1}) with
expected value $1/\sum_{j=1}^d \lambda_j$.  If $k \ge 2$, then for the
second mutation there are $d-1$ choices.  However, the rate at which
the second mutation occurs now depends on the specific realization of
the first mutation.  Hence, we have to consider the set $\mathcal C_k$
of all total orderings
\begin{equation}
    T_{j_1} < \dots < T_{j_k}
\end{equation}
of $k$ out of $d$ waiting times.
%We again define stochastic variables by the time-ordering,
%Eq.~(\ref{eq:10}). Yet in the situations with distinct rates $u_j$,
%the $\tau_j$ may not be expressed by a linear chain because the rates
%of the following mutations depend specifically on which mutations have
%occured before. Hence one must compute the contributions of all
%possible combinations separately.
There are $(d)_k = d!/(d-k!)$ such orders.  We identify $\mathcal C_k$
with the set of all mutational pathways $j_1 \rightarrow \dots
\rightarrow j_k$ of length $k$ in $2^{[d]}$.  For notational
convenience, we write such a path $C \in \mathcal{C}_k$ as a
collection of subsets $C = (C_0, C_1, \dots, C_k)$ such that $C_0 =
\emptyset$ and $C_i = \cup_{\ell=1}^i \{j_\ell\}$, for $i = 2, \dots,
k$. Each set $C_i$ represents an intermediate genotype on the path
with $i$ mutations.

The expected waiting time until any $k$ out of $d$ mutations occur is
the weighted sum over all mutational pathways of length $k$,
\begin{equation}  \label{eq:indep}
    \EE[\tau_k] = \sum_{C\in \mathcal C_k} \EE[\tau_k \mid C] \Prob[C],
\end{equation}
where
\begin{equation}  \label{eq:pathprob}
    \Prob[C]= \prod_{i=1}^k \frac{\lambda_{j_i}}
        {\sum_{j \in \Exit(C_{i-1})} \lambda_{j}}, 
\end{equation}
is the probability of pathway $C$ with $\{j_i\}=C_i\setminus C_{i-1}$
and $\Exit(C_{i-1})=[d] \setminus C_{i-1}$ being the set of all
possible mutations at step $i$. Furthermore, 
\begin{equation}  \label{eq:pathtime}
    \EE[\tau_k \mid C]= \sum_{i =1}^k \frac{1}
        {\sum_{j \in \Exit(C_{i-1})} \lambda_j}
\end{equation}
is the expectation of the waiting time $\tau_k$ given that the path
$C$ is realized.  For a fixed
pathway, say $1 \rightarrow \dots \rightarrow k$, the waiting time
distribution is $\Exp(\lambda_1 + \dots + \lambda_d)$ for the first
mutation,, $\Exp(\lambda_2 + \dots + \lambda_d)$ for the second
mutation, and $\Exp(\lambda_j + \dots + \lambda_d)$ for the $j$-th
mutation.  In general, Eq.~(\ref{eq:pathtime}) arises from a linear
$k$-step process, Eq.~(\ref{eq:tau}), with transition rates $u_j =
\sum_{\ell \in \Exit(C_{j-1})} \lambda_\ell$.  Note that this
waiting time is different from the waiting time in the linear model,
because here a linear pathway is considered within a much larger
lattice of mutational patterns (Figure~\ref{fig:graph}(b)).  In the
denominators of both Eqs.~(\ref{eq:pathprob}) and (\ref{eq:pathtime})
we account for alternative evolutionary routes by summing over the
fixation rates of all mutations that could have occurred at this
point.
% In the following we will denote this set of possible mutations at
% step $i$ by $\Exit(C_{i-1})$.

If all fixation rates are identical to $\lambda$, then $\Prob[C] =
1/(d)_k$ and $\EE[\tau_k \mid C] = \sum_{i=1}^k 1/(d - i + 1) \lambda$ are
independent of $C$, and we recover
%$\EE[\tau_k] = (1/u)\sum_{i=1}^k 1/(d - i+ 1)$
Eq.~(\ref{eq:waitingtimeiid1}).

\subsection{Partially ordered mutations}
\label{sec:ct-cbn}

Sequentially and independently accumulating mutations can be regarded
as two opposite extreme cases, where the linear model imposes maximum
constraints on the order in which mutations can occur, while the
independent model imposes none.  For most biological systems,
including cancer progression, we expect more realistic models to lie
somewhere in between these extremes
(Figure~\ref{fig:graph}(c)). Conjunctive Bayesian networks are a class
of waiting time models that allow for partial orders among the
mutations, i.e., they encode constraints like $T_i < T_j$ for some of
the mutations
\citep{Beerenwinkel2006a,Beerenwinkel2007d,Beerenwinkel2007e}.

Formally, the (continuous time) conjunctive Bayesian network is
defined recursively by
a partially ordered set, or poset, $P = ([d], \prec)$ and
fixation rates $\lambda_j$, as
\begin{equation}  \label{eq:cbn}
    T_j = \bigl\{ \max_{i \in \pa(j)} T_i \bigr\} + \Exp(\lambda_j),
    \qquad j = 1, \dots, d,
\end{equation}
where $\pa(j) = \{ i \mid i \prec j$ and
$(i \prec \ell \prec j \Rightarrow i=\ell$ or $\ell = j) \}$
is the set of mutations that cover mutation $j$.
This model class includes the linear and the independent
model, for which $|\pa(j)| = 1$ and $\pa(j) = \emptyset$,
respectively. It is a Bayesian network model, because the joint
density of $T = (T_1, \dots, T_d)$ factors into
conditional densities as
%\begin{widetext}
  \begin{equation}
    \begin{split}
      f_{T_1,\dots,T_d}(&t_1,\dots,t_d) = \prod_{i=1}^d
      f_{T_i|\{T_j:j\in \pa(i)\}}(t_i\mid \{t_j : j\in\pa(i)\})\\
    &  = \prod_{i=1}^d \lambda_i \exp\bigl( -\lambda_i[t_i-\max_{j\in
          \pa(i)} t_j] \bigr) \II \bigl(t_i > \max_{j\in \pa(i)}
        t_j\bigr).
    \end{split}
  \end{equation}
%\end{widetext}
The expected waiting time until $k$ mutations have accumulated
according to the partial order $P$ can be calculated in a
fashion similar to Eq.~(\ref{eq:indep}).
Let $J(P) \subset 2^{[d]}$ denote the set of all genotypes that
are compatible with the poset $P$, i.e., the subsets $S \subset [d]$
for which $j \in S$ and $i \prec j$ implies $i \in S$.
Considering the set $\mathcal{C}_k(P)$ of all mutational pathways
of length $k$ in $J(P)$, we find
\begin{equation}  \label{eq:cbntime}
    \EE[\tau_k] = \sum_{C \in \mathcal{C}_k(P)} \EE[\tau_k \mid C] \Prob[C],
%     \left( \prod_{i=1}^{k} \frac{\lambda_{j_i}}{\sum_{j\in\Exit(C_{i-1})}\lambda_j }\right)
%     \left( \sum_{i=1}^{k}  \frac{1}{\sum_{j\in\Exit(C_{i-1})}\lambda_j}\right).
\end{equation}
%Here $C=\{C_i\in J(P) \mid C_i\subset C_{i+1} \forall i=0,\ldots d\}$
%denotes a (maximal) chain of mutations in the lattice, $\mathcal{C}(J(P))$ being
%the set of all possible chains.
% where $\lambda_{\Exit(C_i)}:=\sum_{j\in [d] \setminus C_i} \lambda_j$ is the
% sum of the fixation rates of all mutations that can occur next
% in genotype $C_i$.
 with $\Prob[C]$ and $\EE[\tau_k
\mid C]$ defined in Eqs.~\eqref{eq:pathprob} and \eqref{eq:pathtime}, respectively.
% This expected waiting time of is of the same form as the corresponding
% Eqs.~(\ref{eq:indep})--(\ref{eq:pathtime}) for the independence model:
% the first factor of each summand is the probability of choosing the
% specific path $C$ and the second factor is the waiting time of the
% path.
The set of possible paths $\mathcal C_k(P)$ is restricted to
the lattice $J(P)$; hence the set of possible next mutations,
$\Exit(C_i)$, is also constrained to the elements compatible with the
poset $P$. The expected waiting time for $k$ mutations grows with the
number of relations because $\Exit(C_i)$ is the larger, the less
relations exist in the poset. It is therefore maximal in a
totally ordered set, then decreases for a partial order, and is minimal
for an unordered set. The two opposing limit cases of the linear
chain and the independent case represent extrema also in terms of the
expected waiting time.

In practice, the number of mutational pathways can be large, but
the expectation, Eq.~(\ref{eq:cbntime}), can be computed recursively
without the need of enumerating all paths.  The conjunctive Bayesian
network does not only allow for calculating the expected waiting time,
but it has also nice statistical properties.  Both the parameters
$\lambda_j$ and the structure $P$ of the model can be inferred 
efficiently from observed data.  The maximum likelihood estimator for
the parameters is
\begin{equation}
  \label{eq:ml}
  \hat \lambda_j = \frac{M}{\sum_{i=1}^M (t_{ij}-\max_{\ell \in
      \pa(j)}t_{\ell j})},
\end{equation}
where $M$ is the number of observations and the $i$-th observation
$t_{i \cdot}$ is a realization of $T = (T_1, \dots, T_d)$. The
maximum likelihood poset $\hat P$ is the maximal poset
that is compatible with the data.
In other words, $\hat P$ is simply
the poset that contains all compatible relations.

In practice, the occurrence times of mutations, $T_{j}$, may not be
observable, but instead only mutational patterns are available. This
setting gives rise to a censored version of the conjunctive Bayesian
network, in which parameter estimation is still feasible using an
Expectation-Maximization algorithm \citep{Beerenwinkel2007e}.

\section{Population dynamics}
\label{sec:population-dynamics}

In the previous section we have treated genetic progression as an
effective process with steps including both mutation and clonal
expansion that occur at effective rates
$\lambda_j$. We will now dissect these two processes and analyze
models with explicit mutation and proliferation. Let $\mu$ denote the
mutation rate. We assume that each mutation increases fitness by the
same amount $s$ in a multiplicative manner such that the fitness of a
cell with $j$ mutations is $(1 + s)^j$. Before analyzing the system
with both mutation and selection we first discuss this model for
$s=0$. This corresponds to an ensemble of $N$ independently and
identically distributed copies of the waiting time process. For
example, such a situation is found in the colon: It consists of
more than $10^6$ crypts \citep{HumphriesNRC2008}, each of which can
develop an adenoma independently. The model also applies to the case of
selectively neutral mutations in a tissue and we will present
expressions for the waiting time until the first cell has accumulated
a given number of mutations.

\subsection{Independent cell lineages}

If $s=0$, all cells have the same replicative capacity irrespective of
their mutational patterns. Mutations therefore accumulate
independently in a neutral evolutionary process. We can analyze this
process by interpreting the independence model with rates $\lambda_j =
\mu$ as describing the state of a single cell. The population of
genetically heterogeneous, but phenotypically identical cells can then
be regarded as an ensemble of independent cell lineages, each 
evolving according to Eq.~(\ref{eq:indepmodel}). 
In this setting, we are interested in the average time it takes until
the first cell with $k$ mutations appears in a population of size $N$,
i.e., in the expectation of
$\min \{ \tau_k^{(i)} \mid i=1,\dots,N\}$,
where all $\tau_k^{(i)}$ are identical distributed according to
Eq.~(\ref{eq:tau}) with $u_j = (d-j+1)\mu$. 
% under the distribution
%$f(t)=\prod_{i=1}^Nf(t^{(i)})=\prod_{i=1}^N \mu x_{k-1}(t^{(i)})$. Since
%this is a feasible task we may reformulate the problem:

Rather than calculating this expectation, we take a different approach.
Let $\tau_k$ be the waiting time for $k \ll d$ independent mutations,
which is equivalently defined by the linear process with rate $\mu d$,
Eq.~(\ref{eq:waitingtimeiid1}).
In an ensemble of many identical cell lineages the probabilities
$x_j(t) = \Prob[\tau_j < t]$
may be identified with the relative abundances of 
cells with $j$ mutations in the population.
Similarly, $\Prob[\tau_k < t]=\sum_{j \ge k}\Pois(j;\mu dt) $ is the
fraction of cells having at least $k$ mutations. When this fraction
exceeds $1/N$, chances are high that the first cell has accumulated
$k$ mutations. Thus, we define
\begin{equation}  \label{eq:taustar}
    \tau_k^* = \inf \, \{t \ge 0 \mid
         x_k(t) \ge 1/N \}.
\end{equation}
This quantity can also be interpreted as the $(1/N)$-quantile of the
distribution of $\tau_k$.  Using Eq.~(\ref{eq:x_k}), we can find
$\tau_k^*$ by solving
\begin{equation}  \label{eq:solve}
    \frac{1}{N} = \Pois(k;\mu d t)\sum_{j=0}^\infty
        \frac{(\mu d t)^j}{(k+j)_j}.
\end{equation}
for $t$.
Since $N$ is typically very large ($N = 10^6$ to $10^9$ cells),
we are searching for solutions in the regime where the
right hand side of Eq.~(\ref{eq:solve}) is small.
This is the case for $\mu d t \ll k$. Then only the $j=0$ term
of the sum contributes appreciably and we have to solve
% For large $N$, the terms $x_j(t)$, $j> k$, do not contribute to the
% infimum and we may approximate $\tau_k$ by
% \begin{equation} \label{eq:tau_def}
% \tau_k \approx \inf \, \{t \ge0 \mid x_k(t) \ge 1/N \}
% \end{equation}
$1/N = \Pois(k;\mu d \tau_k^*)$.

For $k=1$, we consider the subset of 1-cells, i.e., cells containing
one mutation, that starts growing in the background of mutation-free
cells as $x_1(t) = \mu d t\exp(-\mu d t) \approx \mu d t$ for $t \approx
0$.  Thus the average waiting time to the appearance of the first cell
with one mutation is $\tau_1^* \approx 1/(\mu d N)$.  Similarly, for
$k=2$, we find $x_2(t) = (1/2) (\mu d t)^2 \exp(-\mu d t) \approx (1/2)
(\mu d t)^2$ and thus $x_2(t) = 1/N$ has the approximate solution
$\tau_2^* \approx \sqrt{2} / (\mu d \sqrt{N})$.  Alternatively, one
can arrive at this approximation by considering the initial linear
growth of the population of 1-cells. The first 2-cell is
produced by these growing 1-cells when
\begin{equation}  \label{eq:first2cell}
    \mu d\, \int_0^{\tau_2^*} x_1(t) \rmd t = \frac{1}{N},
\end{equation}
having the same approximate solution given
above.

In general, the solution of $1/N = \Pois(k;\mu d \tau_k^*)$
is given in terms of the Lambert $W$ function,
which is defined as the solution of $W(z) e^{W(z)} = z$, 
\begin{equation}
  \label{eq:tau*_exact}
  \tau_k^* = - \frac{k}{\mu d} \, W_0 \left( -
    \frac{k!^{1/k}}{k \, N^{1/k}} \right),
\end{equation}
where $W_0$ is the principle branch of $W$
\citep{Corless1996}.
For large population sizes $N$, the argument of the
Lambert $W$ function in Eq.\ (\ref{eq:tau*_exact}) is close to zero and hence
$W(z) \approx z$. We obtain
\begin{equation}
  \label{eq:tau*_approx}
  \tau_k^* \approx  \frac{k!^{1/k}}{\mu d \, N^{1/k}}, \qquad \mbox{for all } k \ge 1,
\end{equation}
which generalizes the approximations for $\tau_1^*$ and
$\tau_2^*$ given above.

\begin{figure} \centering
\includegraphics[width=\linewidth,trim=0 0 0 40]{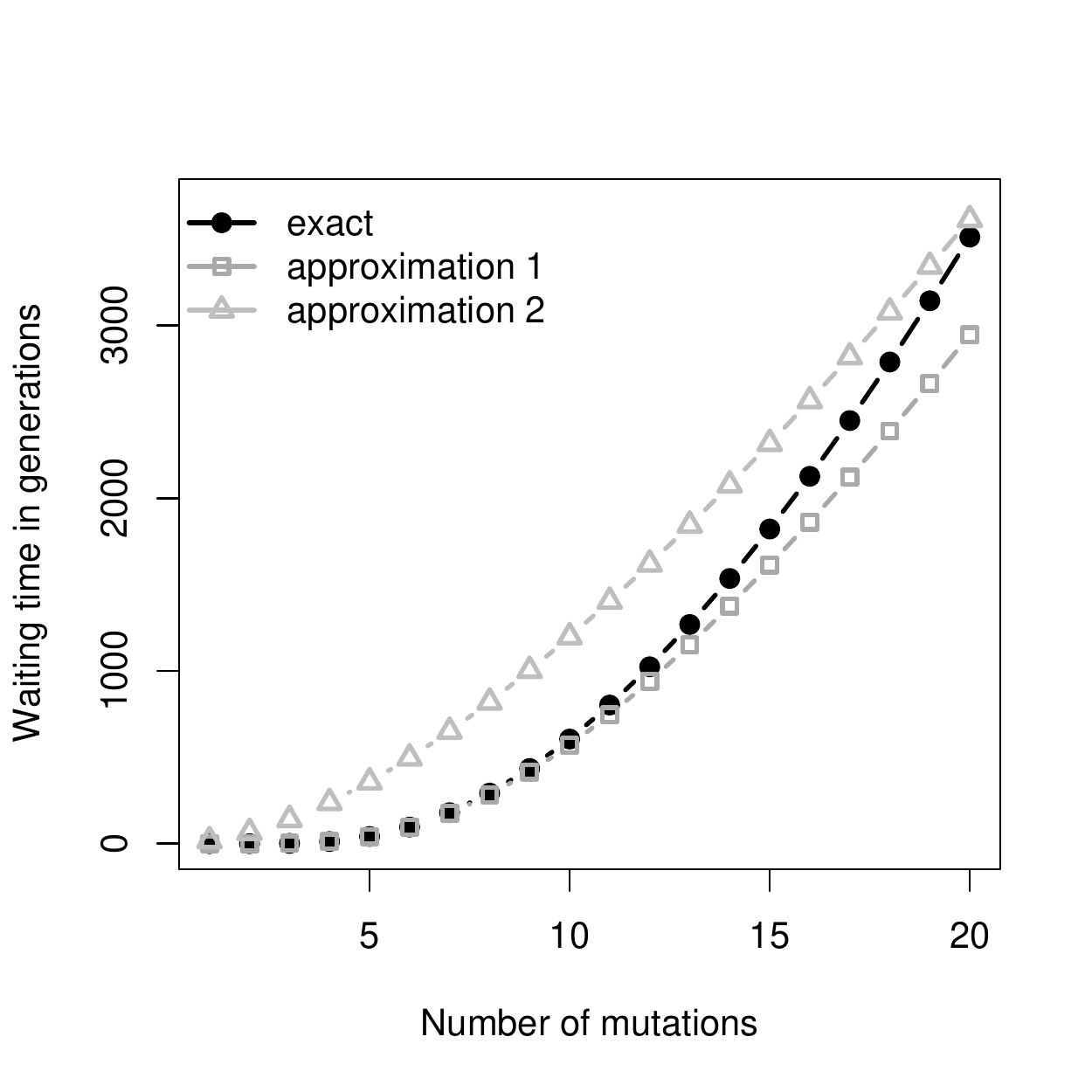}
\caption{\label{fig:tau} 
    Approximate solutions of the waiting time $\tau_k^*$
    as defined by the equation $1/N = \Pois(k;\mu d \tau_k^*)$,
    for $N=10^9$, $\mu d = 0.001$, and $k=1,\dots,20$. 
    The exact solution in terms of the Lambert $W$ function
    (filled circles, Eq.~(\ref{eq:tau*_exact})) is compared to 
    the approximation given in Eq.~(\ref{eq:tau*_approx})
    (squares) and the less accurate but simpler
    approximation of Eq.~(\ref{eq:tau_neutral}) (triangles).
}
\end{figure}

On the other hand, for large $k$, we have
$ k!^{1/k} \approx k/e$ and
%\begin{equation}
$ N^{1/k} \approx 1 + (\log N) / k$
%\end{equation}
leading to
\begin{equation}
  \tau_k^* \approx \frac{k^2}{e \mu d (k + \log N)}.
\label{eq:tau_neutral}
\end{equation}
This approximation is less accurate, but reasonable for usual
parameter values and $k=0,\dots,20$ (Figure~\ref{fig:tau}). For $N=1$,
it coincides with the result for a single cell line,
Eq.~(\ref{eq:waitingtimeiid1}), up to a constant factor of $1/e$.  The
waiting time depends on the the inverse of the logarithm of the
population size. For example, the average waiting time to the first
cell with $k$ mutations among $10^9$ cells is only about 20 times
shorter than the same waiting time in a single cell. For large $k$,
this expression becomes again linear in $k$ (Figure~\ref{fig:tau}).

The normal mutation rate due to DNA polymerase errors is on the order
of $10^{-10}$ to $10^{-9}$ base pairs (bp) per cell per generation
\citep{KunkelARB2000}. For an average human gene size of 27kbp
\citep{VenterS2001}, the mutation rate per gene should be on the order
of $\mu\approx 10^{-6}$ per cell per generation. The waiting time until the
first of $10^9$ cells has accumulated $k= 20$ mutations would be
on the order of $10^{6}$ cell generations which, in turn, typically
occur at the time-scale of days or weeks. Thus, the waiting time would be on
the order of $10^6$ days or more, clearly exceeding a
human lifetime. Hence a neutral evolutionary process alone cannot account
for the genetic progression of cancer.

\subsection{Selection and clonal expansion}

We now analyze the dynamics of an evolving cell population in which each
mutation confers the same selective advantage $s > 0$. Because a new
mutant with an additional mutation has a growth advantage, it will
expand in the tissue and outcompete the other cells.  The next
mutation is most likely to occur on this growing clone.  We therefore
use an evolutionary model of carcinogenesis that accounts for mutation
and selection \citep{Beerenwinkel2007c} and trace the number of cells
with $j$ mutations, $N_j(t)$, in each generation $t=0,1,2,\dots$.

\begin{figure}[t]
  \centering
\newcommand{\A}{*+[o][F]{~0~}}%{{\bigodot}}
\newcommand{\B}{*+[o][F]{~1~}}%{{\bigotimes}}
\newcommand{\C}{*+[o][F]{~2~}}%{{\bigotimes}}
\[\xymatrix@=0.08\linewidth{
\A\ar[d] &\A\ar@{~>}[d] &\A & \A\ar[d]\ar[dl] &\A &\A\ar[d]\ar[dl] &
t= 0\\
\A &\B\ar[dl]\ar[d] \ar[dr]  &\A &\A\ar[d]  &\A\ar@{~>}[d]  &\A\ar[d] & t= 1 \\
\B &\B\ar[dl]\ar[d]  &\B \ar@{~>}[d]\ar[dr] &\A &\B \ar[d] \ar[dr] &\A & t= 2\\
\B &\B &\C &\B &\B &\B & t= 3\\
}\]
\caption{Wright-Fisher process. 
    Illustrated are four generations of a single realization
    of the Wright-Fisher process with population size
    $N = 6$.  In each generation $t+1$,
    cells are drawn randomly from the previous generation $t$
    according to the multinomial distribution 
    given in Eq.~\eqref{eq:multinomial}.
    Directed edges ($\rightarrow$) indicate the genealogy of this realization.
    In general, cells with more mutations are more likely to generate
    offspring and will therefore, on average, outcompete
    cells with fewer mutations. In each generation,
    cells are subject to mutation ($\rightsquigarrow$).
    In this realization, the waiting time to the first 
    appearance of a cell with $k=2$ mutations was $\tau_2 = 3$
    generations.
}
  \label{fig:Wright-Fisher}
\end{figure}
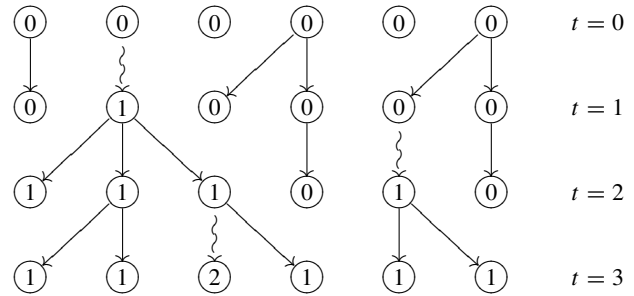

Consider a population of $N$ cells that undergo subsequent rounds of
cell divisions as shown in Figure~\ref{fig:Wright-Fisher}. In each
generation, mutations occur randomly and independently at rate $\mu$.
The total number of possible mutations is denoted $d$. We assume that
fitness, i.e., the expected number of offspring, is proportional to
$(1+s)^j$, where $j$ is the number of accumulated mutations.  The
population dynamics are assumed to follow a Wright-Fisher process
\citep{Ewens2004}.  In this model, generations are time-discrete and
synchronized.  A new configuration $[N_0(t+1),\ldots, N_d(t+1)]$ of
cells is drawn from the previous generation $t$ according to the
multinomial distribution
  \begin{equation} \label{eq:multinomial}
    \begin{split}
      \Prob \left[ N_0(t+1) = n_0, \dots, N_d(t+1) = n_d \right]\\ = 
      \frac{(n_0 + \dots + n_d)!}{n_0!\cdots n_d!}  \prod_{i=1}^d
      \theta^{n_j}_j,
    \end{split}
  \end{equation}
where $n_0 + \dots + n_k = N$. The parameters $\theta_j$
denote the probability of sampling a $j$-cell,
\begin{equation}  \label{eq:theta}
    \theta_j =
    \sum_{i=0}^j\binom{d-i}{j-i}\mu^{j-i}(1-\mu)^{d-j}
    \frac{(1+s)^{i}x_i}{\sum_l(1+s)^lx_l},
\end{equation}
where we defined $x_j(t) = N_j(t)/N$ as the relative abundance
of $j$-cells. A cell with $j$ mutations can occur in generation
$t+1$ either as progeny of a $j$-cell in generation $t$,
or by erroneous duplication of a $(j-1)$-cell.
For $s=0$ and infinitesimal generation times, the model
reduces to the case of independent cell lineages
undergoing independent
mutations, which has been discussed in the previous section.

%This model has recently been proposed by  and
%the authors have shown the existence of traveling clonal waves and
%derived formulae for the waiting times in this evolutionary setting.

In general, no closed form solution of the Wright-Fisher process
is known. However, the dynamics defined by
Eqs.~(\ref{eq:multinomial}) and (\ref{eq:theta}) display
certain regularities that can be exploited in order to derive
an approximate analytical expression for the expected
waiting time to the first cell with $k$ mutations, $\tau_k^*$.
Numerical simulations show that the subsets of $j$-cells
sequentially sweep through the population
and the mutant waves travel at constant speed
\citep{Beerenwinkel2007c}.
This regular behavior can be analyzed by decomposing the process
into the generation of a new cell type by mutation and
its clonal expansion driven by selection.

The dynamics of clonal expansions 
are given by the replicator equation \citep{Nowak2006},
\begin{equation}  \label{eq:replicator}
    \dot{x}_j(t) = s x_j(t) \left[ j-\sum_{i=1}^\infty
        i \, x_i(t) \right],
\end{equation}
where we consider only those cell types that are already present in
the system and we ignore mutation.  The fitness of $j$-cells is
$(1+s)^j \approx 1 + js$, if $s \ll 1$.  Eq.~(\ref{eq:replicator}) has
a solution in terms of the Gaussians $x_j(t)=A
\exp(-[j-vt]^2/[2\sigma^2] )$ with normalization constant $A$ and
width $\sigma^2=v/s$, where $v$ is the velocity of the traveling wave.
The initial growth of a newly founded clone is exponential, but
eventually follows this Gaussian distribution. The final decline
corresponds to the clone ultimately becoming extinct by outcompetition
of fitter clones harboring additional mutations.

The velocity of the waves is determined by the mutation process.  A
new $(j+1)$-cell is generated by mutation from the growing clone of
$j$-cells. The equation $x_{j+1}(t) = 1/N$ can therefore be
rewritten, similar to Eq.~(\ref{eq:first2cell}), as
\begin{equation}  \label{eq:velocity}
    \int_0^{\tau_{j+1}^*} \mu d x_{j}(t) \rmd t = \frac{1}{N},
\end{equation}
where initially, $x_{j}$ grows exponentially according to Eq.~\eqref{eq:replicator}.
%Note, however, that $x_j(t)$ now also includes the
%reproduction dynamics.
This approach finally yields the approximate expected waiting time
\citep{Beerenwinkel2007c}
\begin{equation}  \label{eq:tauWrightFisher}
    \tau_k^* \approx
    \frac{k \log^2 (s / [\mu d])}
    {2s \log N}.
\end{equation}
This expression suggests approximating the Wright Fisher process,
Eqs.~(\ref{eq:multinomial}) and (\ref{eq:theta}), by a linear
multistep process, Eq.~(\ref{eq:tau}), with transition rate $u = (2s
\log N) / \log^2 (s/[\mu d])$ in which stages correspond to clonal
expansions \citep{MaleyCL2007}.  Comparing
Eq.~\eqref{eq:tauWrightFisher} with the waiting time in a neutral
evolutionary process, Eq.~(\ref{eq:tau_neutral}), here the waiting
time per mutation, $1/du$, contributes only logarithmically and the
expected waiting time Eq.~(\ref{eq:tauWrightFisher}) is proportional
to $k/s$, reducing the overall waiting time considerably.  The reason
for this acceleration lies in the growth advantage of the mutated
cells: A single cell produces an exponentially growing number of
clonal offspring. This growth, in turn, directly relates to the
probability of creating a cell with an additional mutation. Therefore,
clonal expansions dramatically speed up the accumulation of mutations
in a population.

For example, considering a fitness advantage of $s=10^{-2}$ per
mutation, $d=100$ susceptible loci, a mutation rate of $u=10^{-7}$ per
gene, and a population size of $N=10^9$ cells results in a waiting
time of $\tau_{20}^* \approx 10^3$ generations.  With a generation
time of 1 to 2 days, this waiting time ranges on the time scale of
several years, being consistent with clinical observations. By
contrast, the waiting time in the neutral model is on the order of
$10^{6}$ generations. Hence even a moderate selective advantage
decreases the waiting time by three orders of magnitude.

The time $\tau^*_j$ denotes the time after which the probability that
a cell with $j$ mutations has been generated exceeds $1/N$. This is an
approximation for the expected waiting time of the first $j$-cell with
an additional mutation in a population of size $N$. For the
Wright-Fisher process it is known, however, that due to genetic drift
the probability of fixation of a selectively advantageous mutation
initially present in a single cell is only $2s$
\citep{Ewens2004}. Hence, the majority of mutated cells become
extinct. This is also observed in the numerical simulations of the
Wright-Fisher process (Eqs.~(\ref{eq:multinomial}, \ref{eq:theta});
\citet{Beerenwinkel2007c}). On average it takes $1/2s$ cells until the
first successful mutant is generated. This effect is included
indirectly in approximation Eq.~(\ref{eq:tauWrightFisher}): $x_j(t)
\propto e^{st}$ is the expected frequency conditioned only on
$x_j(0)=1/N$ and not on survival. It also accounts for all
trajectories, including those going extinct.
% Replacing $1/N$ with $1/2sN$ in Eq.~(\ref{eq:velocity}), the
% fixation probability $2s$ enters the logarithm on the
% right-hand-side of Eq.~(\ref{eq:tauWrightFisher}) which then reads
% $\log(s/\mu d /2s) $. For typical parameter values only one out of
% fifty mutants reaches fixation. However, this prolonges the waiting
% time $\tau^*_k$ only by a factor of 2. Hence genetic drift does not
% significantly slow down the progression of cancer, because new
% mutants keep being generated at increasing speeds from the growing
% clones.
Recently, this effect has been studied in a related model
 \citep{DesaiG2007, BrunetG2008}.
\citet{DesaiG2007} found an approximate waiting time of
$\tau_k^*\approx k
\log(s/[ud])/(s[2\log N + \log\{sud\}])$. Comparing with expression
Eq.~\eqref{eq:tauWrightFisher}, the only difference
is the term $\log(sud)$ in the denominator. For typical parameter
values, $\log(sud) \approx -7 \approx -\log N$. Therefore, the waiting
time is larger by a factor of $1.5$ as compared to
Eq.~\eqref{eq:tauWrightFisher}. But this comparison is
limited, because the models are not identical. For
example, \citet{DesaiG2007} obtain a fixation probability of $s$,
whereas in the Wright-Fisher model it is $2s$.

\section{Conclusion}
A quantitative understanding of cancer progression is required for
constructing clinical markers and for revealing rate-limiting steps of
this process. Here, we have analyzed waiting time models for
carcinogenesis and solved the equations defining the expected waiting
times. Similar quantities have previously been shown to measure the
degree of tumor progression and to predict survival in cancer patients
\citep{Rahnenfuehrer2005}.

In the simplest case, carcinogenesis may be described by a linear
multistep process.
%, where each step denotes a specific transition. 
The progression stages are generally described by histological
alterations and functional changes, or on a molecular level, by 
mutation of certain genes and subsequent clonal expansion. In a
general multistep process, the overall waiting time to reach stage $k$
is the sum of the waiting times of all predecessing steps.

If tumor stages are defined by the number of mutations that have
fixated in the cell population, then the progression dynamics depends on
the order in which mutations accumulate.  For example, mutations may
accumulate in a linear fashion, according to a partial order, or completely independently.  Linear
accumulation is the slowest and independent progression is the
fastest. The acceleration can be considerable, especially if many
mutations are available that drive carcinogenesis.

The linear and the independent model present opposing limits of the
conjunctive Bayesian network family of models, in which the mutations
obey a partial order. The relations of the poset may result from
causal relationships among  mutations, such as the requirement in
colon cancer for the tumor suppressor {\em APC} to be mutated before
other mutations are beneficial and can fixate.  The poset constraints
induce a subset of mutational pathways in the hypercube representing
all combinatorial genotypes (Figure~\ref{fig:graph}). Because the
waiting time is additive in the mutational pathways, its expected
value for the conjunctive Bayesian networks ranges between those
for the linear and the independent model.

% In general, more than $1\%$ of human genes
% are associated with cancer \citep{FutrealNRC2004}; each tumor,
% however, typically harbors only $\sim 100$ mutations, of which
% $\sim20$ are thought to be relevant \citep{Sjoeblom2006}. Among these,
% some are required before other mutations may exploit their malignant
% potential. A typical example is the mutation of the {\it APC} gene
% after which a subsequent mutation of the {\it KRAS2} gene results in a
% clonal expansion \citep{Vogelstein2004}.

%This model class also  allows an efficient
%estimation of the underlying poset and the model parameters in an EM
%algorithm \citep{Beerenwinkel2007e}. This, in turn provides insights
%into the biological mechanisms of tumorigenesis.

We have discussed a particular instance of the
Wright-Fisher process, an evolutionary model comprising mutation and
selection. In this model, we find waiting times on the order of 20
years for a normal mutation rate and a selective advantage of 1\% per
mutation.  The successive clonal waves might be regarded as the stages
in classical multistage theory.
A neutral evolutionary process can not explain the clinical
progression of colon cancer, in which about 20 out of hundreds of
mutations accumulate in a time frame of 5 to 20 years.  This process
may only be explained by advantageous mutations giving rise to
clonal expansions.  These selective sweeps drastically increase the
chances of acquiring additional mutations the spreading
offspring.

\bibliographystyle{mybib}
%\bibliography{d:/User/nikob/lit/nikos,lit}

\bibliography{nikos,lit,lit2}
%\bibliography{nikos,mps-bib}

%\newpage
%
%\begin{appendix}
%\section{Expectation maximization algorithm}
%
%Mathematical details of ...
%
%\end{appendix}

\end{document}